%% file: main.tex
\documentclass{llncs}

\usepackage[english]{babel}
\usepackage[utf8]{inputenc}
\usepackage{latexsym,amssymb,amstext,amsfonts,amsmath,graphicx,setspace,mathtools,bm}
\usepackage[colorinlistoftodos]{todonotes}
\usepackage{enumerate} 
\usepackage{paralist}
\usepackage{float}
\usepackage{array}
\usepackage{lipsum}

\usepackage{tikz}
\usetikzlibrary{fit,positioning}
\usetikzlibrary{calc}

\begin{document}

\title{A Restaurant Process Mixture Model for Connectivity Based Parcellation of the Cortex}
\titlerunning{Cortical Connectivity Based Parcellation}  % abbreviated title (for running head)
\author{Daniel Moyer\and Boris A. Gutman \and Neda Jahanshad\and \\ Paul M. Thompson}
%\author{--}
%
\authorrunning{Daniel Moyer et al.} % abbreviated author list (for running head)
%\authorrunning{--} % abbreviated author list (for running head)
\tocauthor{Daniel Moyer, Boris Gutman, Neda Jahanshad, and Paul Thompson}
%\tocauthor{--}
%
\institute{Imaging Genetics Center, University of Southern California\\
\email{moyerd@usc.edu}
%\institute{--\\
}

%\author{Anonymous}

%\institute{}

\maketitle              % typeset the title of the contribution

\begin{abstract}
One of the primary objectives of human brain mapping is the division of the cortical surface into functionally distinct regions, i.e. parcellation.
While it is generally agreed that at macro-scale different regions of the cortex have different functions, the exact number and configuration of these regions is not known.
Methods for the discovery of these regions are thus important, particularly as the volume of available information grows. Towards this end, we present a parcellation method based on a Bayesian non-parametric mixture model of cortical connectivity.
\end{abstract}

\input{intro.tex}

\input{model/model.tex}

\input{results/results.tex}
\input{disc/disc.tex}

\subsection*{Acknowledgements}

This work was supported by NIH Grant U54 EB020403, as well as the NSF Graduate Research Fellowship Program. The authors would like to thank the reviewers as well as Greg Ver Steeg for multiple helpful conversations. 

\bibliographystyle{splncs03}
\bibliography{cbp}

\end{document}

%% file: intro.tex
\section{Introduction}

%Towards this end, we present a connectivity based parcellation method which takes into account the structure of the set of pairwise interactions between the regions it estimates, without requiring a specified number of regions.

Historically, researchers proposed and investigated regional brain parcellations through manual dissection and qualitative description \cite{zilles2010centenary}. % of a small number of post-mortem specimens, usually at a cyto-architectural scale 
The rise of non-invasive neuro-imaging coupled with advances in computing and computer vision allowed for the exploration of automated parcellation methods, both for fitting existing atlases to data and for data-driven discovery of functionally and structurally cohesive parcels \cite{van2012parcellations}. The success of the former propelled the rise in interest and analyses of brain connectomics in the last decade \cite{sporns2005human}. Connectomics is a topic of interest within all scales of neuroscience; at the macro-scale, it is often defined by discrete networks of cortical gray matter regions as nodes with weighted or binary edges connecting them. In structural connectomics--the focus of this paper--the edge weights are usually based on counts of estimated structural connections recovered using diffusion MRI tractography, sometimes weighted by a microstructural measure.

A number of papers have focused on the converse, using the connection profile of either voxels or vertices, or in the case of functional MRI, the pairwise signal correlation for each vertex pair to define the parcellation of the cortex (see \cite{eickhoff2015connectivity} and the references therein).
%This immediately brings up a problem: if our representation of connectivity is based on the parcels, then how can we learn parcels from our representation of connectivity? This circularity persists even outside of later statistical analyses of resulting networks.\NJ{ $\longleftarrow$ I don't think this argument makes sense given the citation, the parcels representing connectivity and the parcels learned from connectivity are on different scales... voxel/vertex and cluster .... so it works and is not that circular. Maybe just remove the last sentence, you talk about how it makes sense next anyway... } 
For connectivity based parcellation (CBP) methods using structural connectivity, two modeling choices are required: 1) a spatial resolution of the grid on which connections are defined, and 2) the criteria on which clusters should be formed. In almost every existing method, connectivity measures are first estimated for a high resolution grid of atomic units (choice one), e.g. voxels or vertices, and the atomic units then combined under spatial constraints optimizing a desiderata (choice two).

The first decision is essentially a division of connectivity into two scales: the macro-region level and higher resolution voxel/vertex level, where our task is to learn from the former the regions of the latter.
The second modeling choice is the criteria on which these atoms are clustered. Many popular choices come from more general clustering literature, e.g. within-group sum of squares (explained variance), within-group statistical distances, and mixture model likelihoods \cite{eickhoff2015connectivity,parisot2015tractography,yeo2011organization}. These criteria use the connectivity profiles of each meso-scale atom without regard to the network structure they induce at the macro-scale. In other words, they treat each vertex or voxel as a data point with an associated vector (its row in the meso-scale adjacency matrix), and then cluster based on this vector space. If one vertex is changed from one group to another, these methods generally do not re-evaluate the quality of all the groups, though on the macro-scale each of their connective profiles would have changed. %While there is growing work examining the effects of these choices individually, little work has been conducted to address both model choices simultaneously. \NewText{we believe there is a more excellent way}.

In the present work, we will address both these choices. We present a method framed in the context of generative models, specifically Bayesian non-parametric mixture models which place priors over all possible partitions of the higher resolution grid, and do not require the number of clusters to be predefined. One large classes of such priors are the so-called ``restaurant processes'', used here. We implement a continuum form of connectivity for our mixture components, and further leverage a conjugate likelihood-prior relationship to produce closed form marginal likelihoods for network interactions, allowing efficient sampling.

Our paper is organized as follows: we first define terminology, rigorously define the parcellation task, and describe the model as a whole. We then describe each of its components in closer detail. We then present results on two datasets, and discuss the model in relation to existing models and methods.

%% file: model/model.tex
\section{Model}
\label{sec:model}

Let $\Omega$ be the white matter/gray matter interface (the inner cortical surface), with the acknowledgment that $\Omega$ is in general composed of two disjoint sheets, each with a boundary at the medial wall. Fix a coordinate system over $\Omega$, and define a parcellation $P$ as any set of regions $\{E_i\}$ where $E_i \subset \Omega$,$\bigcup E_i = \Omega$, and $|\bigcap E_i | = 0$ (i.e. the regions $E_i$ are almost disjoint). We assume there exists a latent parcellation $P^*$ that accurately describes the cortical surface with respect to its underlying neuroanatomical structure. Our objective is the recovery of $P^*$, specifically using structural connectivity information, and without specifying the exact number of regions. In order to accomplish this, we construct a joint generative model of parcellations and connectivity.

%Towards this end, we utilize two \NewText{technologies}, namely the distance dependent Chinese Restaurant Process (ddCRP) \cite{blei2011distance}, and Poisson-Gamma conjugation, modified for the Poisson process.

We start by choosing a model of partitions. We use the distance dependent Chinese Restaurant Process (ddCRP) \cite{blei2011distance}, a variant of the popular Chinese Restaurant Process (CRP) non-parametric Bayesian models. CRP models are most commonly used in mixture models, providing a prior over all possible label assignments for any number of label-parameter pairs. A main assumption of the CRP is the exchangeability of the data; the ddCRP removes this exchangeability assumption, allowing for non-trivial topologies of dependence between data points. This is discussed in Section \ref{sec:ddcrp} in detail. Briefly, ddCRP allows us to use non-parametric style mixture models on mesh grids, where we assume \textit{a priori} that neighboring patches are dependent. For example, we assume there is spatial auto-correlation over the discrete manifold of the mesh. Practically speaking, the ddCRP is the component responsible for merging or splitting the parcels (clusters), and in general for their configuration.

We next choose a mixture component model; the distribution chosen here will generate the observed network between estimated regions from the ddCRP. We choose to follow the style of the Infinite Relational Model \cite{kemp2006learning,baldassano2015parcellating}, where we model interactions between clusters instead of the profiles of the clusters themselves. Thus, we need a separate parameter for each \emph{pair} of regions. Before diving into this however, it is important to consider the form of our connectivity data. Structural connectivity is estimated using streamlines (tractography), usually via identifying tracts which intersect the cortical surface at two locations. Thus, the evidence of connectivity is a set of endpoint pairs on $\Omega$.

In traditional connectivity analysis, these endpoints are counted by region pair, and a graph is formed from the resulting count statistics. These representations abstract away both knowledge of region geometry such as surface area, curvature etc, as well as topological information, i.e. region adjacency; this is the information we will be using in the ddCRP model. While it is possible to ignore these conflicting motives and directly kluge a graph to a spatial patch model, we instead attempt to retain spatial intuition in our connectivity representation.

%\NewText{In this way, our problem is motivated by preserving both geometric and streamline-based connectivity in our parcellation.} While it is possible to ignore these conflicting motives and directly kluge a graph to a spatial patch model, we instead attempt to retain spatial intuition in our connectivity representation.

Consider $\Omega\times\Omega$, the set of all possible tract endpoints intersecting the cortical surface. We model the observation of these pairs of endpoints as a spatial point process on $\Omega\times\Omega$. Assuming that each tract is independently recovered\footnote{It is important to make the distinction between physical fascicles and recovered tracts. Here, we define the latter to be the reconstructed tractography.}, this process is the Poisson point process. That is, for any region pair $E_i \times E_j \subset \Omega\times\Omega$, the number of tract endpoint pairs observed in that region pair is Poisson distributed with parameter $\int_{E_i}\int_{E_j}\lambda(x,y)dydx$. Here $\lambda:\Omega\times\Omega\rightarrow\mathbb{R}^+$ is a non-negative rate function assumed to be integrable over all $\Omega\times\Omega$. For Poisson processes, $\lambda$ completely characterizes the process.
While we discuss further the Poisson point process in Section \ref{sec:pois}, in the view of the overall model it is important to note one convenient property: disjoint regions have independent counts. 

Moving back to the mixture components, we make the following simplifying assumption on the form of the tract endpoint process: each region pair interacts in a homogenous manner. That is, we assume $\lambda$ is constant over any pair of parcels. Thus, for any finite configuration of $K$ parcels we have on order $K^2$ non-negative scalar parameters to estimate, and these parameters are the rate parameters for Poisson spatial processes generating the evidence of connectivity. We choose to use the Gamma distribution to model these parameters (i.e., each pair of parcels $(g_i,g_j)$ draws a rate $\lambda_{ij}$ from a Gamma prior). As shown in Section \ref{sec:pois}, the conjugacy of the Gamma distribution with the homogenous Poisson process allows for closed form marginal distributions, and thus efficient collapsed sampling methods. We also choose to use the mesh faces $\{f_m\}_{m=1}^M$ as the elements of our ddCRP-mixture. This is because connectivity is usually defined over intersections of tracts with areal units, and both the ddCRP as well as the Poisson process naturally operate over such regions.

Putting this all together, and leaving the meaning of $c_{m}$ for the next section, the model is as follows:
\begin{align*}
c_{m}, g_{i} & \sim \text{ddCRP}(\alpha, Adj)\\
\lambda_{ij} & \sim \text{Gamma}(a,b)\\
D_{ij} & \sim \text{Poisson Point Process} (\lambda_{ij})
\end{align*}

\subsection{The distance dependent Chinese Restaurant Process (ddCRP)}

\label{sec:ddcrp}

%Both of these (and, in general, the family of so-called Restaurant Processes) are distributions over partitions; \NewText{in our context, these will be partitions of $\Omega$, i.e. possible parcellations}.

As suggested by its name, distance directed Chinese Restaurant Process is a variant of the Chinese Restaurant Process (CRP), often used in non-parametric Bayesian mixture models as a prior over possible mixture components (i.e. a distribution over distributions). Let $\alpha$ be some positive constant concentration parameter, and let $G_0$ be a prior distribution over mixture component parameters (for us, a gamma distribution). The original CRP mixture model \cite{pitman2002combinatorial} describes an endless stream of customers (data) entering a restaurant with an infinite number of tables (clusters). Each customer either chooses (with prescribed probability dependent on $\alpha$) to sit at an existing table (which has a particular component distribution) or sit an unoccupied table (draw a new component distribution from the prior). Up to the indexing of the tables, for any finite number of observations any number of clusters and configuration of cluster associations is possible.

In the original CRP, the data are assumed to be exchangeable; that is, the joint likelihood of any observations is invariant under permutations of observation indices. However, in our spatial context we have a topology of face adjacencies. Permutations of the face indexes are non-trivial, and thus the faces are not exchangeable. %Instead, for data with meaningful dependencies (in this case, topological), Blei and Frazier \cite{blei2011distance} introduce the ddCRP variant.
To model this, the ddCRP allows each customer to choose another customer (possibly itself) to sit with based on its dependencies. This forms a directed graph of seating choices; table assignments are then made to each group of customers who have chosen to sit with each other, i.e. each connected component of the seating choice graph. Mixture components are drawn for each table from $G_0$, and only then are the actual data drawn from each mixture component. Clearly this is a two stage procedure. In our context, this means each face will choose to be in a cluster with one of its neighbors or itself.

As above, let $\{f_m\}_{m=1}^M$ be the set of mesh faces, and let $\{c_m\}_{m=1}^M$ be the corresponding assignments, where each $c_m \in \{1,\dots,M\}$. We draw $c_m$ conditioned adjacency information $Adj$ as follows:

\[p(c_m = j | Adj) \propto \left\{ \begin{array}{cc} 1 & \text{if }j\text{ is adjacent to }i\\
\alpha & \text{if } m = j\\
0 & \text{otherwise}\end{array}\right.\]

We denote each cluster of faces as $g_k$ for $k \in \{1,\dots,K\}$, where $K$ is the number of clusters. Due to our restriction of $c_m$ to the indices of faces adjacent to $f_m$, each $g_k$ is a contiguous region, and the set of groups forms a valid parcellation of $\Omega$. We note that the original ddCRP is defined for more general distance functions.

%Note that the original ddCRP is defined more generally for distance functions.\BlueComment{Maybe we'll explain why we don't?}

\subsection{Mixture components: Poisson-Gamma}
\label{sec:pois}

The evidence of pairwise interaction between regions in structural connectivity is the set of tract endpoints $D = \{(x_t, y_t)\}_{t=1}^T$. Since the regional clusters are defined over discrete grids of areal atoms (mesh faces), these are naturally aggregated to count measures over each pair of sub-regions. For any pair of regions $(g_i, g_j) \subset \Omega \times \Omega$, define $D_{ij} = \{(x_t,y_t)  \in g_i\times g_j\}$. We model the counts $|D_{ij}|$ using the Poisson process with fixed intensity $\lambda_{ij}$, where the area $g_i \times g_j$ contains a random count $|D_{ij}|$ distributed
\[|D_{ij}| \sim Poisson( \int_{g_i\times g_j} \lambda dxdy )\]
Using the independence assumption of the tract endpoints, the likelihood of any configuration of tract endpoints can then be written
\[\mathcal{L}( D ) = \prod_{g_i,g_j} \exp\left\{-\int_{g_i\times g_j} \lambda_{ij} dxdy\right\}\lambda_{ij}^{|D_{ij}|} \]
We use a Gamma prior for the $\lambda_{ij}$ parameters, the conjugate prior of the Poisson distribution. Using the Gamma distribution allows us derive a simple closed form marginal distribution for $D_{ij}$ that ``integrates out'' the $\lambda_{ij}$'s, leaving a likelihood in terms of prior parameters $a,b$. It is as follows:
\begin{align*}
P(D_{ij} | a,b) & = \int P(D_{ij},\mu| a,b) d\mu = \int P(D_{ij} | \mu ) P(\mu | a,b) d\mu \\
& = \int \underbrace{\exp\left\{-\int_{g_i}\int_{g_j}\lambda dA\right\} \prod_{t=1}^{|D_{ij}|} \lambda}_{\text{Homogeneous Point Process}}\times \underbrace{\frac{b^{a}}{\Gamma(a)}\exp(-b\lambda)\lambda^{a-1}}_{\text{Gamma Prior} }d\lambda\\
& = Z(a,b)\int \underbrace{\exp\left\{ -(|g_i \times g_j| + b)\lambda\right\} \lambda^{|D_{ij}| + a - 1}}_{\text{Un-normalized Gamma Posterior}}d\lambda\\
& = \frac{Z(a,b)}{Z(a',b')} = \left(\frac{\beta}{|g_i \times g_j | + b}\right)^{a}\left(\frac{1}{|g_i \times g_j | + b}\right)^{|D_{ij}|}\frac{\Gamma(a + |D_{ij}|)}{\Gamma(a)}
\end{align*}
Here, $Z(a,b) = \frac{b^{a}}{\Gamma(a)}$
% We make the following simplifying assumptions: first, that there exists a ``correct'' parcellation $P^*$ such that its regions divide the cortical surface into its functional modules \BlueComment{find a new word for this }, and, second, that between each pair of regions $E_i,E_j$ tracts are observed according to a homogenous (constant) rate $\lambda(E_i,E_j)$.
\subsection{Combined Model and Collapsed Sampling Scheme}

%Putting the ddCRP together with the Poisson-Gamma mixture components, we have a model for surface region association from tract endpoint data. The model has the following form:

%\begin{align*}
%c_{m}, g_{i} & \sim \text{ddCRP}(\alpha, Adj)\\
%\lambda_{ij} & \sim \text{Gamma}(a,b)\\
%D_{ij} & \sim \text{Homogenous Poisson Point Process} (\lambda_{ij})
%\end{align*}

We will estimate the model via Collapsed Gibbs Sampling, specifically using the closed form integral over $\lambda_{ij}$ to avoid sampling the interaction parameters. Starting at iteration $\ell=0$, we update each $c_m^{\ell}$ by the following conditional likelihood:
\begin{align*}
P(c_m^{\ell + 1} = k | D ) & \propto P(c_m = k) \prod_{i,j} P(D_{ij} | a,b, c_m^{\ell + 1} = k, \{c_{r}^{\ell + 1}\}_{r < m}, \{c_{s}^{\ell}\}_{s > m})
\end{align*}
Since we assume $c_m$ is restricted by our mesh topology, we have only a small number of options to evaluate. We denote the seating graph edge $c_m^{\ell}$ as a ``critical edge'' if for any other node $f_{m'}$ such that $c_{m'} = m$ there exists a path to the face with index $c^{\ell}_{m}$. Let $g_{old}$ be $f_m$'s previous component, and $g_{crit}$ be the component of $f_{m}$ without its own edge (its critical component). Without loss of generality we may further order each neighbor of $f_{m}$ as $f_{n}$ for $n = 1,\dots,N$, and their groups as $g(n)$.
%Further, in general some components will be unaffected by any of the possible values $c_m$ can take; these can be ignored, as they are constant.
Using these definitions, for triangle meshes we can write out all possible scenarios:
\begin{enumerate}
\item \label{non-crit-sur} If $c_m$ is not critical, and all neighbors are of the same component before the update, then we can simply choose $c_m$ via $P(c_m = k)$, as there is no difference with respect to the induced components.
\item \label{non-crit-non-sur} If $c_m$ is not critical, but not surrounded by the same component, then we are asking essentially ``Should $c_m$'s previously induced component join one of its as of yet independent neighbors''. Thus, 
\begin{align*}
P(c_m^{\ell + 1} = n | D ) & \propto P(c_m = n) \prod_{k}^K P(D_{*,k} | g_{new} = g_{old} \cup g_{n} , g_{k})\\ &~~~ \times \prod_{\substack{g_{\hat{n}}: \hat{n}\neq n \\ \hat{n} \neq old}}^N \prod_{k}^K P(D_{g(n),k} | g_{\hat{n}}, g_{k})\end{align*}
for each neighbor $n$. Here $D_{*,k} = \{(x_t,y_t) \in g_{old} \cup g_{n} \times g_{k} \}$
\item \label{crit} If $c_m$ is critical, then for each neighboring component (including the component $g_{old} \backslash g_{crit}$ in the neighbors) we have
\begin{align*}
P(c_m^{\ell + 1} = n | D ) & \propto P(c_m = n) \prod_{k}^K P(D_{*,k} | g_{new} = g_{crit} \cup g_{n} , g_{k})\\
&~~~ \times \prod_{\hat{n}\neq n }\prod_{k}^K P(D_{\hat{n},k} | g_{\hat{n}},g_k).
\end{align*}
\end{enumerate}

We iteratively update the face associates using $P(c_m^{\ell + 1} = k | D )$, collecting samples after every pass. While this generates a posterior distribution over $c_m$, we simply take the maximum \textit{a posteriori} (MAP) estimate as our selected parcellation.

In general, updates made in Gibbs sampling algorithms are done sequentially; this is because strong dependencies between concurrent updates will destabilize some samplers. However, in cases of low dependence approximate asynchronous parallel updates have been used with empirically strong results (so called ``Hogwild'' updates \cite{johnson2013analyzing}). In our case, most updates are either within components, or between small components (with correspondingly small interdependencies), so a small degree of parallelism is possible. In practice we use a compromise between the serial algorithm and the parallel version: we use a shared memory parallel sampler for calculating the likelihoods of a small batch $c_i$, then make a serial updates based on these likelihoods. This allows a roughly linear speed-up in the number of threads used, though there is a slowly scaling cost of the serial update.

\subsection{Implementation notes}

In fixing a coordinate system, it is common to split the white matter/gray matter interface into two spheres, each with a null region where the corpus callosum bridges the longitudinal fissure. Thus, an easy system can be constructed using spherical coordinates and a marker for hemisphere.

The symmetry of each tract's endpoints requires careful consideration to avoid double counting; while the intuition of the model can be understood without thinking about the symmetry of the data, when evaluating joint probabilities it is important to only include each data point once. This can be achieved by only evaluating $P(D_{ij} | a,b)$ for $i \leq j$. When computing the parallel updates, in our experience it is much more efficient to keep the threads active but idle, and simply have a single thread do the serial update. This avoids the overhead of repeated thread spawns, which for some implementations/architectures can be costly.

%% file: results/results.tex
\section{Procedure and Results}
\label{sec:results}

In order to test our proposed model, we use two open datasets, one composed of 20 subjects each scanned twice from the Institute of Psychology, Chinese Academy of Sciences (IPCAS) subset of the Consortium for Reliability and Reproducibility (CoRR) dataset \cite{zuo2014open}, and the other composed of 30 subjects from the Human Connectome Project (HCP) S900 release \cite{van2013wu}. The pre-processing differs slightly between the datasets, to account for the different imaging parameters. In general the HCP dataset has higher resolution (both in voxel size and angular resolution) leading to different tractographies. On each dataset we compare the performance of the proposed method against two recommended alternatives: Ward's method, a greedy hierarchical clustering method \cite{eickhoff2015connectivity}, and Spectral Clustering \cite{parisot2015tractography}. 

\subsection{Preprocessing and Tractography}

\textbf{IPCAS}: T1-weighted (T1w) and diffusion weighted (DWI) images were obtained on 3T Siemens TrioTim by the original investigators \cite{zuo2014open} using an 8-channel head coil and 60 directions. Each subject was scanned twice, roughly two weeks apart. T1w images were processed with Freesufer's \cite{fischl2012freesurfer} recon-all pipeline to obtain a triangle mesh of the grey-white matter boundary registered to a shared spherical space \cite{fischl1999high}. We resample this space to a geodesic grid (where each face has approximately equal area) with 10,000 total faces, doing so only after computing tract intersections with the surface. Probabilistic streamline tractography was conducted using the DWI in 2mm isotropic MNI 152 space, using Dipy’s \cite{garyfallidis2014dipy} implementation of constrained spherical deconvolution (CSD) \cite{tournier2008resolving} with a harmonic order of 6.  Tractography streamlines were seeded at 2 random locations in each white matter voxel labeled by FSL's FAST. Streamline tracking followed directions randomly in proportion to the orientation function at each sample point at 0.5mm steps, starting bidirectionaly from each seed point with 8 restarts per seed. As per Dipy’s Anatomically Constrained Tractography (ACT) \cite{smith2012anatomically}, we retained only tracts longer than 5mm with endpoints in likely gray matter.

\textbf{HCP:} We used the minimally preprocessed T1-weighted (T1w) and diffusion weighted (DWI) images rigidly aligned to MNI space. Briefly, the preprocessing of these images included motion correction and eddy current correction (DWI), and linear and nonlinear alignment (betweek T1w and DWI).  We used the HCP Pipeline (version 3.13.1) FreeSurfer protocol to run an optimized version of the recon-all pipeline that computes surface meshes in a higher resolution (0.7mm isotropic) space. We again resample this space to an geodesic grid after computing tract intersections with the surface. 
Tractography was conducted using the DWI in the native 1.25mm isotropic voxel size in MNI space. Probabilistic streamline tractography was performed as in IPCAS above.

\subsection{Fitting and Results}

We fit the proposed method using our parallel sampling scheme, using 60 passes of the sampler with 8 parallel threads (approximately 600,000 updates per subject), using $a=1$, $b=1$, and $\alpha=0.01$. We use the MAP estimate as our results. We fitted Ward clustering by maximizing Explained Variance over a na\"{i}ve search of every possible merge. For the Spectral Clustering method we use an exponential kernel, using the normalized cosine distance as a metric. We use a number of eigenvectors equal to the number of clusters. For both baselines we take the vector of connections as our feature vector. In both clustering schemes, we specify the number of clusters to be equal to that of the proposed method.

\begin{figure*}[t]
\includegraphics[width=0.48\textwidth]{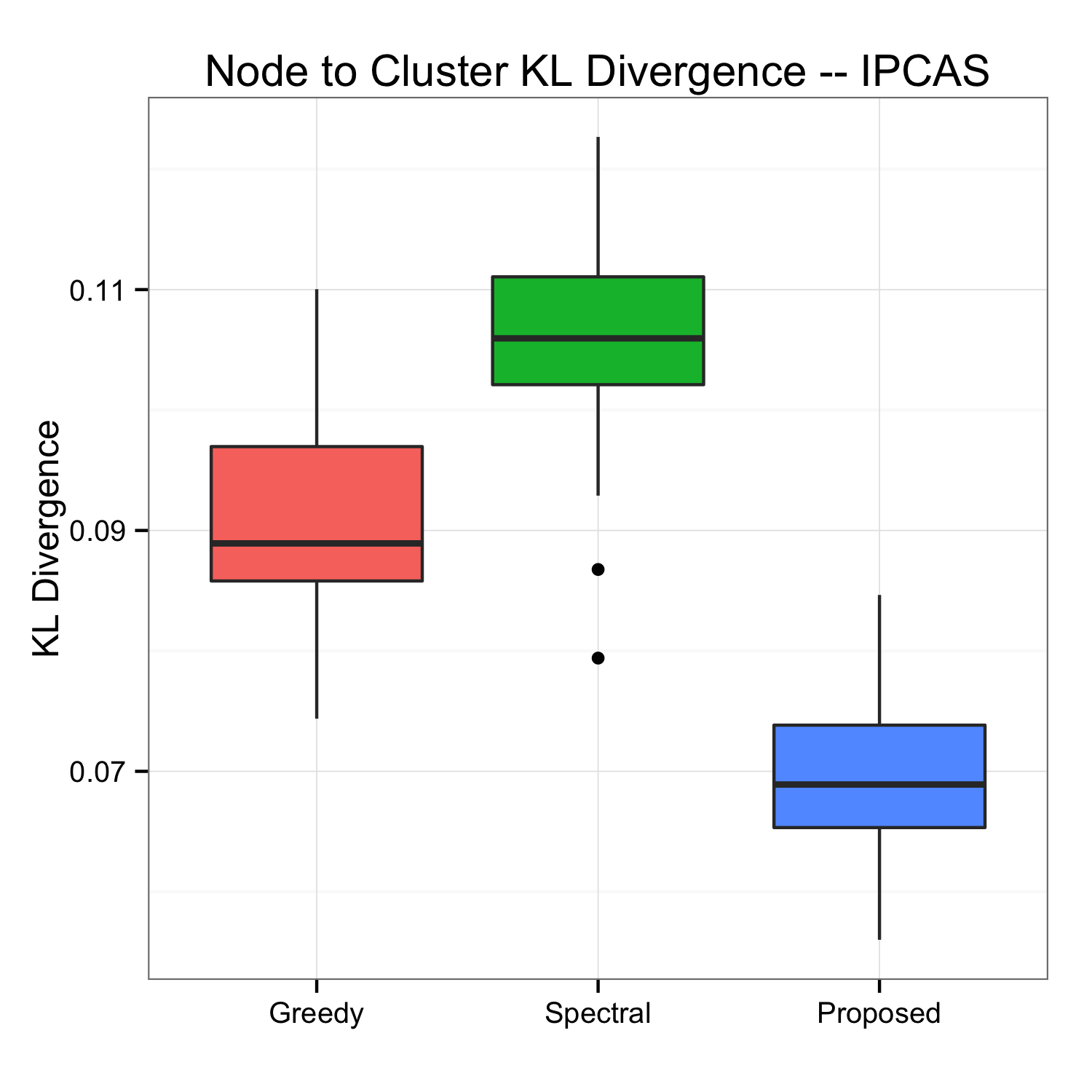}
\includegraphics[width=0.48\textwidth]{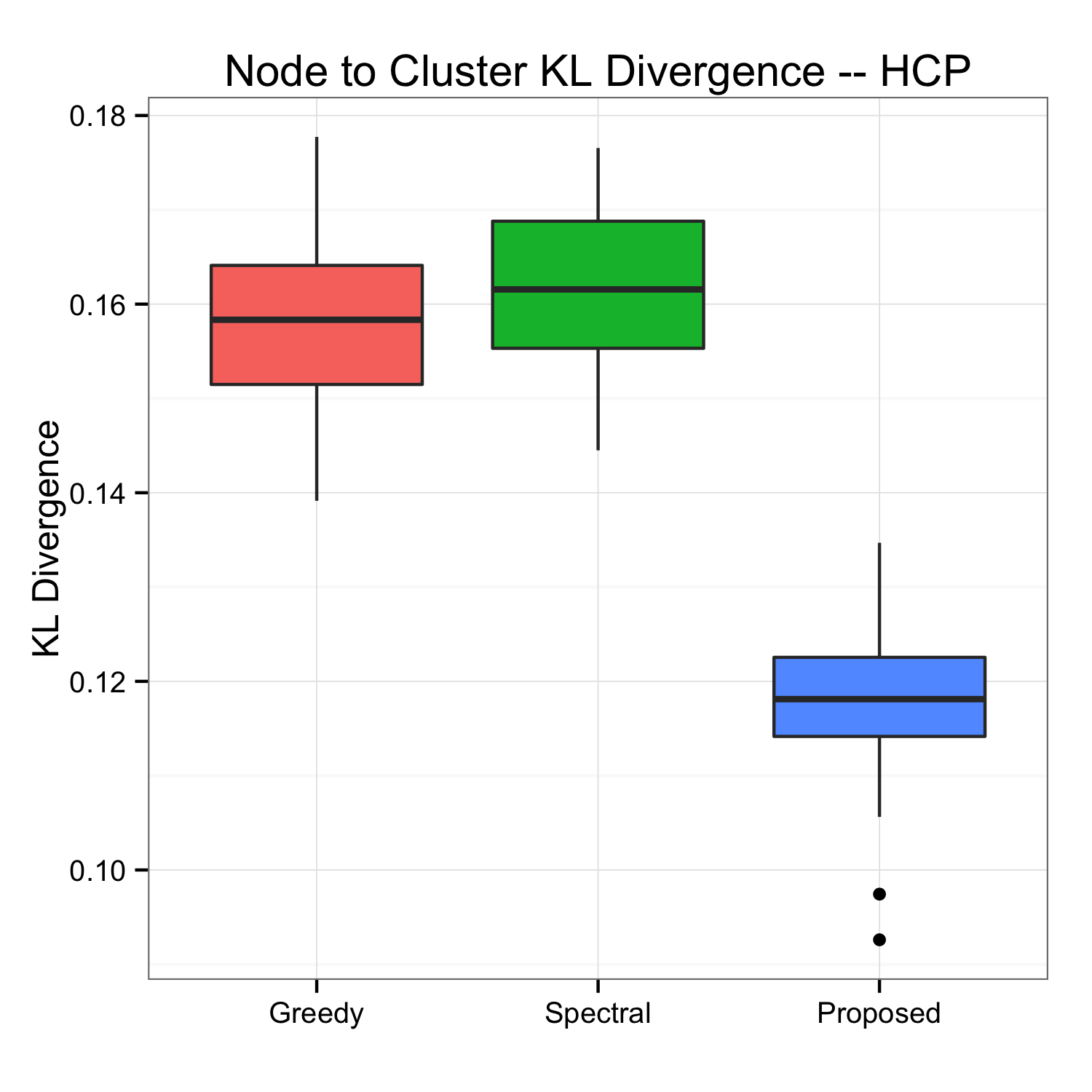}
\caption{Plots of the KL Divergence based goodness of fit measure, for the three methods, on both datasets. Here, \textbf{lower} is better.}
\label{fig:hkl}
\end{figure*}

\begin{figure*}[t]
\includegraphics[width=0.48\textwidth]{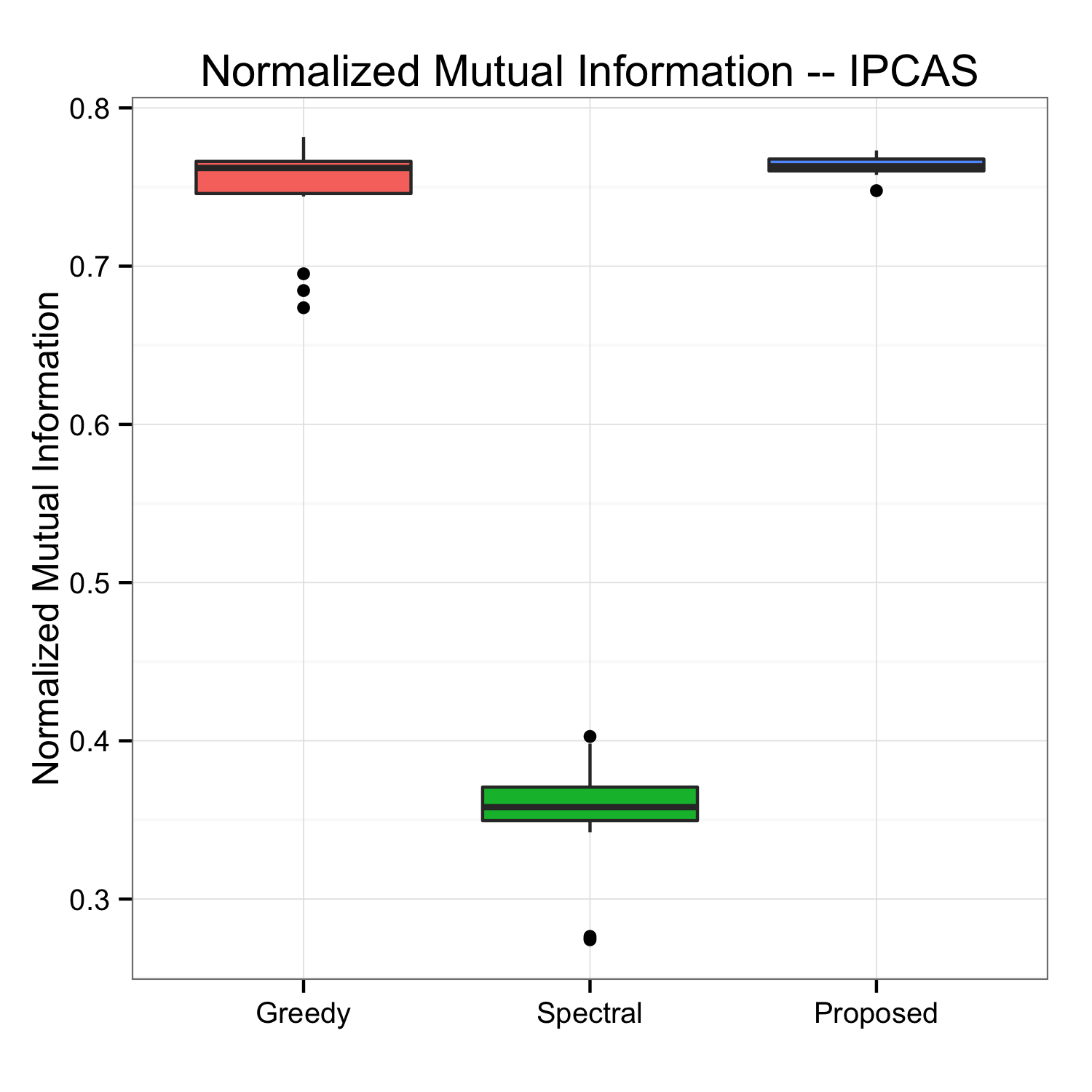}
\includegraphics[width=0.48\textwidth]{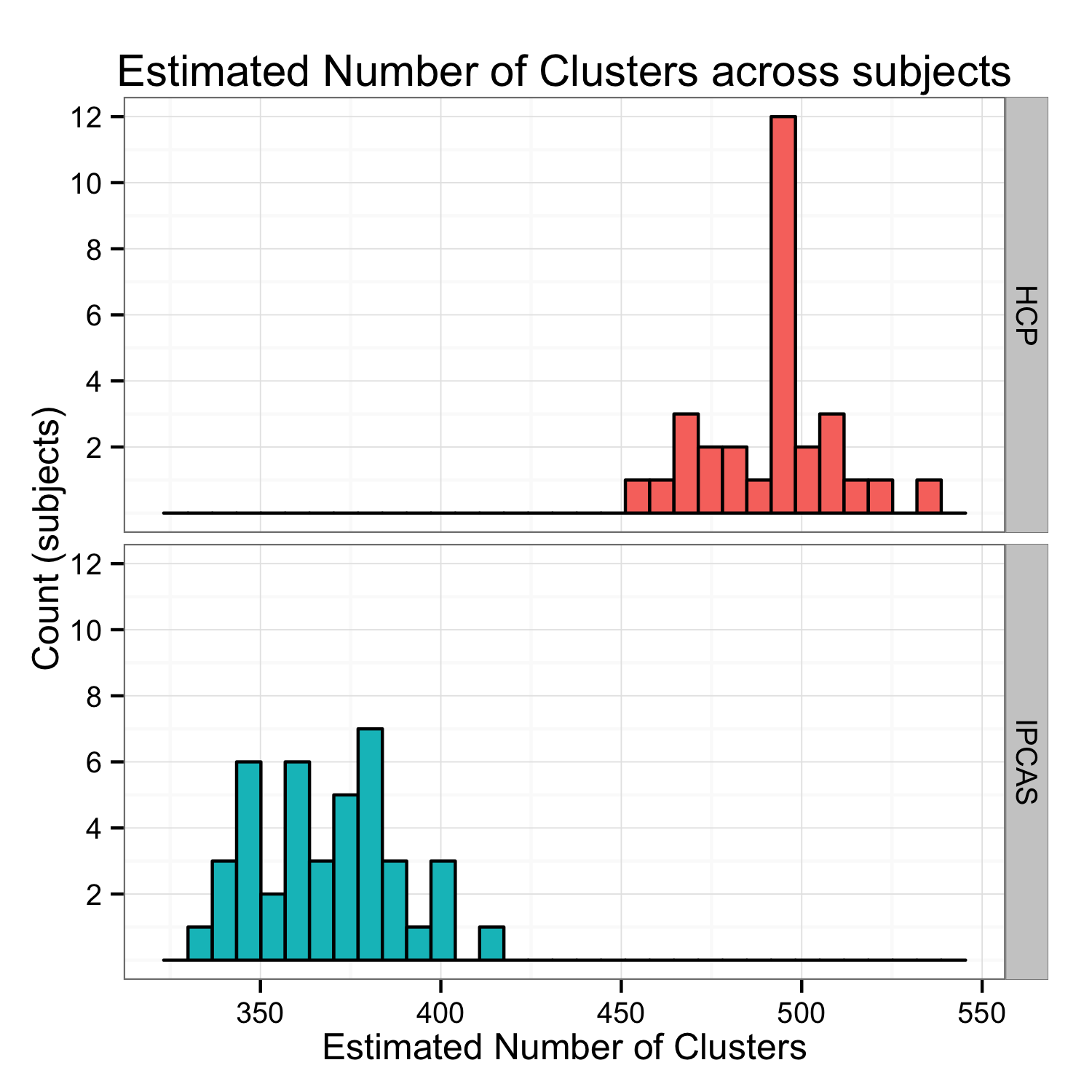}
\caption{Left: Normalized Mutual Information between Test-Retest scans. Here, \textbf{higher} is better. Right: histograms of the number of clusters selected for each subject.}
\label{fig:nmi}
\end{figure*}

We assess cluster quality using a KL-divergence based measure. We take the number of tracts from each face $f_m$ to each region $g_i$ as the objective distribution, and measure how well this is approximated by the average number of tracts from $g(f_m)$ to $g_i$. These form two matrices of dimension $M\times K$. We then normalize these matrices to sum to one and measure their KL divergence as in \cite{parisot2015tractography}. If a cluster is well represented by its average connectivity profile, then this divergence will be low. For the IPCAS dataset we have an additional measure of Test-Retest reproducibility. This is measured by Normalized Mutual Information (NMI)\cite{eickhoff2015connectivity,baldassano2015parcellating}, which measures cluster similarity without requiring similar numbers of clusters. Let $Z$ be a binary matrix of cluster assignments, where, for each row $i$, each entry $Z_{ij}$ is $1$ if $f_{i}$ is in cluster $j$ and zero otherwise. NMI is defined as $I(Z_1,Z_2)/\sqrt{(H(Z_1)H(Z_2)}$, where $I(\cdot,\cdot)$ is mutual information, $H(\cdot)$ is entropy, and $Z_1$ and $Z_2$ are the cluster assignments for the first and second scan respectively. (This uses the convention customary in information theory that $0\log 0 = 0$). NMI is also invariant under permutations of labels. As can be seen in Figure \ref{fig:hkl} and Figure \ref{fig:nmi}, the proposed method is performing well compared to the baseline methods. HCP dataset uses more clusters (around 250 per hemisphere) than the IPCAS dataset (around 175-200 per hemisphere). The difference here may be due in part to the higher resolution of the HCP dataset, leading to greater resolving power with respect to the regional connections. These averages are at the upper range of the number suggested by Van Essen et al. \cite{van2012parcellations}.

%Let $Z$ be binary matrix of cluster associations, where, for each row $i$, each entry $Z_{ij}$ is $1$ if $f_{i}$ is in cluster $j$ and zero otherwise. Let $A$ be the matrix of counts of tract endpoints between each pair of faces. Z has dimension $M \times K$ and $A$ has dimension $M \times M$. Their product $AZ$ is then the count of the number of tracts from each face to each region. Let $C = Z^TAZ$, the usual $K\times K$ connectome adjacency matrix for regions. Let $\tilde{C}$ be the adjacency normalized by the product surface area of each region. The product $ZC$ is the approximation .

%We would like to note that these results are \emph{not} definitive since there are a very large number of variants and tuning parameters for each method (including our own); however, we do feel this suggests that our method is valid.

\begin{figure*}[h]
\includegraphics[width=\textwidth]{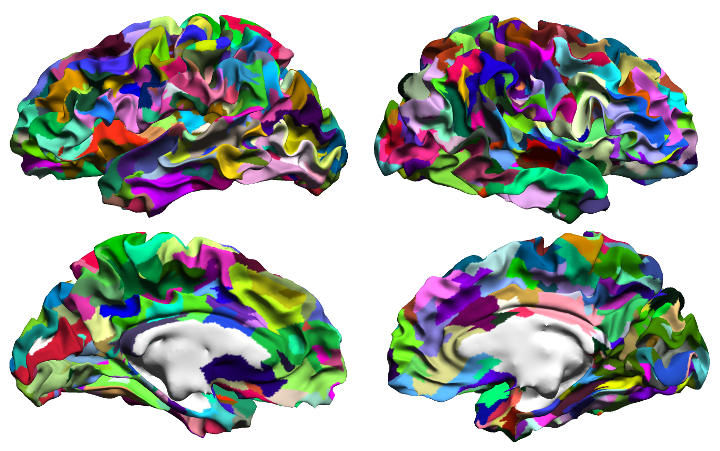}
\caption{An exemplar parcellation from an HCP subject. Region colors are random.}
\label{fig:merged}
\end{figure*}

%% file: disc/disc.tex
\section{Discussion}
\label{sec:disc}

This model draws on the wide range of previously proposed methods in connectivity based parcellation. Several non-parametric Bayesian methods have been proposed, in particular two excellent works Jbabdi et al. \cite{jbabdi2009multiple} and Baldassano et al. \cite{baldassano2015parcellating}, both of whom use Normal-inverse-Wilshart conjugations as their mixture components (Baldassano et al. use a special case, the Normal-inverse-$\chi^2$). These models also enjoy closed form marginal distributions, but do not have infinite divisibility (the distributions they model are not spatial processes). Jbabdi et al., whose work predates the ddCRP, use a Dirichlet Process with spatial priors as their partition prior. They then further define a hierarchical process on top of this that links multiple subjects. Baldassano et al. use the ddCRP directly, but model voxel connections, again without the aid of a spatial process. Instead, they model the aggregate connectivity as coming from a normal distribution.

The ddCRP is similar to a Markov Random Field model with a very strong spatial prior. These models have been successful in obtaining parcellations from functional connectivity \cite{honnorat2015grasp,ryali2013parcellation}, though few if any have used Bayesian non-parametrics. This frame of reference leads us toward more traditional computer vision tasks such as pixel labeling, where as in many cases surface parcellation has been framed as vertex parcellation \cite{clarkson2010framework,parisot2015tractography,yeo2011organization}. This is a small but relatively important conceptual difference; the pixel and mesh-face models have areal units, but vertex parcellations are graphs of infinitesimal points. The intuition of the former leads us toward the use of spatial processes.

A similar spatial process viewpoint of connectivity is proposed in Moyer et al. \cite{moyer2016continuous}, but the discovery of new parcellations is not discussed. Poisson count processes for network interactions have also been explored in the literature \cite{moyer2015mixed}, as have infinite relational variants \cite{hinne2015probabilistic} though usually in the context of network clustering via the stochastic blockmodel (i.e. clustering the regions themselves). These usually ignore spatial constraints.

Alternative methods to Bayesian models usually specify the number of clusters. Of note is Parisot et al. \cite{parisot2015tractography} and a subsequent work by the same authors \cite{Parisot2016}, which propose spectral methods for the parcellation task, augmented with a pre-processing local agglomeration. These papers note the propensity for Spectral Clustering to form equi-areal clusters; as can be seen in Figure \ref{fig:merged}, our method does not form equi-areal groups. Thus, it may be the case that a lower number of clusters for spectral clustering may perform better.

As there is a rich body of functional and anatomical knowledge regarding the cortex, parcellations based on connectivity information alone would need proper neuroanatomical, histological and functional validation, and more information from these sources would ideally be used to optimize parcellations. The model presented here uses only spatial constraints and connectivity to estimate feasible parcellations based on recoverable structural connections from imaging. However, we believe that the modeling techniques explored here can easily be imputed into larger, multi-modal models, and in general the improvements made may increase the accuracy and reproducibility of studies of connectivity patterns. These are critical to furthering our understanding of the living human brain.

%will allow for reproducible studies that are critcally needed to better understand the topography of the living human brain.
%improving the ability to more reliably parcellate the cortical structure using the explicit modeling techniques implemented here provides significant improvements to the use of non-invasive MRI and diffusion MRI metrics in more accurate and precise identifications of connections patterns and improve the sensitivity of misconnections, and deviations from normal patterns, which are being actively studied as biomarkers for disease and disease progression. Improved connectomics methods such as that presented here will allow for reproducible studies that are critcally needed to better understand the topography of the living human brain.